\documentclass[aps,prl,reprint,superscriptaddress]{revtex4-1}
\usepackage{graphicx}
\usepackage{amsmath}

\draft 

\begin{document}

\title{Dispersion of coupled mode-gap cavities} 


\author{Jin Lian,$^{1,*}$ Sergei Sokolov,$^1$ Emre Y\"uce, $^1$ Sylvain Combri\'e,$^{2}$  Alfredo De Rossi,$^{2}$ and Allard P. Mosk }
\affiliation{Complex Photonic Systems (COPS), MESA+ Institute for
Nanotechnology, University of Twente, P.O. Box 217, \\AE 7500  Enschede, The Netherlands \\
$^2$Thales Research and Technology, Route Départementale 128, \\91767 Palaiseau, France\\
$^*$Corresponding author: j.lian@utwente.nl
}

\makeatletter
\let\setrefcountdefault\relax
\makeatother
\date{\today}

\begin{abstract}
The dispersion of a coupled resonator optical waveguide (CROW) made of photonic crystal mode-gap cavities is pronouncedly asymmetric. This asymmetry cannot be explained by the standard tight binding model. We show that the fundamental cause of the asymmetric dispersion is the inherent dispersive cavity mode profile, i.e., the mode wave function depends on the driving frequency, not the eigenfrequency. This occurs because the photonic crystal cavity resonances do not form a complete set. We formulate a dispersive mode coupling model that accurately describes the asymmetric dispersion without introducing any new free parameters.\end{abstract}


\maketitle 

\noindent In coupled cavity systems in photonic  crystals (PhC), light transport occurs via evanescent field coupling between high-quality factor (Q) cavities. The well known example is the coupled resonator optical waveguide (CROW) which is a linear chain of cavities \cite{Yariv1999}. Arrays of coupled cavities have attracted substantial scientific attention for practical applications such as slow light  engineering and strong light-matter interaction enhancement \cite{Xia2007,Hartmann2008,Yanik2004}, and many novel phenomena of fundamental interest such as gauge fields \citep{Hafezi2013,UmucalIlar2011} and time-reversal of light pulses \citep{Longhi2007,Yanik20042nd}.

Realizations of low loss and compact CROWs require high-Q, wavelength-sized cavities. Mode-gap cavities \cite{Kuramochi2006} created by shifting some of the holes of around PhC waveguides as shown in Fig. 1(a) have been demonstrated to be extremely suitable for creating large scale cavity arrays \cite{Notomi2008,Matsuda2014}.   

The tight binding (TB) model \citep{Yariv1999,Ashcroft} is the usual approach for modeling the dispersion in coupled cavities \cite{Chak2006,Sumetsky03,Hartmann2008,Yanik2004}. The core concept of the TB model is that in coupled cavities the wave functions are tightly confined in each  cavity. The eigenmodes of the individual cavities are then coupled to yield waveguide modes. Coupling between neighboring cavities is due to the overlap of the eigen wave function of the cavity modes. The dispersion of the CROW band as predicted by the TB model is a cosine curve \cite{Yariv1999}. For a given structure, its parameters can be evaluated by numerical methods, such as plane wave expansion (PWE) \cite{Leung1990} and finite difference time domain (FDTD) \cite{Taflove2005}. 
\begin{figure}[htp]
\centerline{\includegraphics[width=0.72\columnwidth]{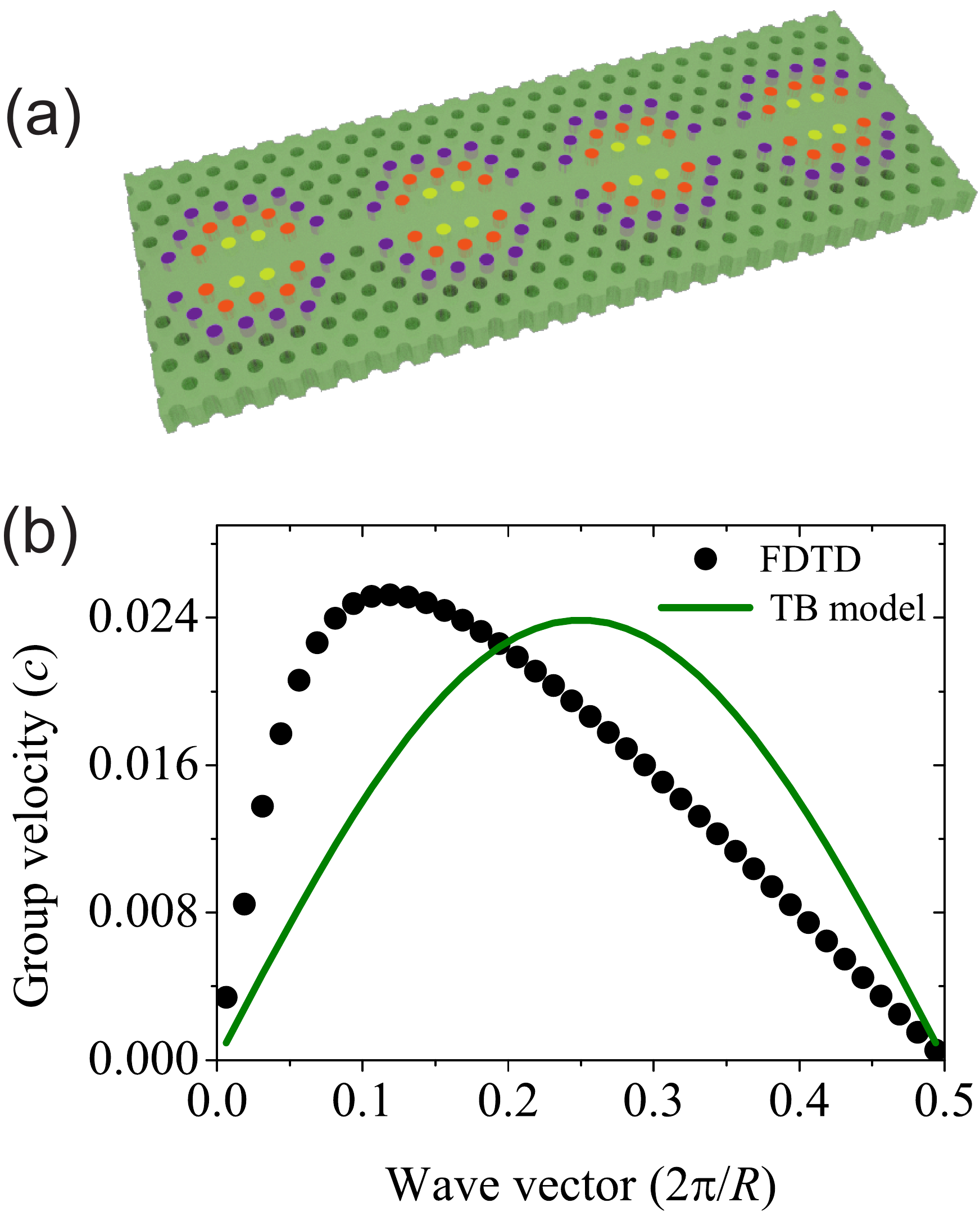}}
\caption{(a) Schematic of a CROW composed of coupled mode-gap cavities. The cavities are created by shifting of the yellow, red and purple holes around the  waveguide in a tapered way. The width of the waveguide is 0.98$\sqrt{3} a$ ($a$ is the lattice constant). The radius of the holes is 0.25 $a$. The shift of the yellow holes is S1=$\eta$ 0.0124$\sqrt{3} a$, $\eta$ is a factor we call it modulation strength. The shifts of the red and purple holes are 2/3 S1 and 1/3 S1 respectively.(b) Group velocity of the 2D CROW consisting of mode gap cavities with modulation strength 1.5 calculated by 2D FDTD (black dots) and TB model (green line).  Parameters are calculated by FDTD.}
\end{figure}

However, the TB model does not  describe the dispersion of a CROW composed of mode-gap cavities correctly. In Fig. 1(b) we show the group velocity curve as obtained from the TB model and as calculated by FDTD \cite{Oskooi2010}. The TB result is perfectly symmetric with respect to $k=0.5 \pi/R$. In contrast, the FDTD result is pronouncedly asymmetric. The  asymmetric spectrum has also been observed experimentally \citep{Matsuda2014}. It is remarkable that the TB model fails here, as it only depends on the assumption of energy-independent eigenmodes, which is a direct consequence of the completeness of the set of eigenmodes \cite{Sakurai}. 

In this Letter,  we uncover the physical origin of the discrepancy and propose an improved mode coupling model.

The resonant modes of a closed conservative cavity form a complete set. An important consequence of completeness is the fact that the response of a cavity is defined in terms of modes do not themselves depend on the driving frequency. In an open cavity we have quasimodes which form a complete set only if certain conditions are fulfilled \citep{Ching1998}, essentially, the edge of the resonant mode must be clearly defined in the structure and no outgoing waves should be scattered back to the cavity. These conditions are never fulfilled in a PhC cavity. In a PhC cavity light is confined by constructive interference of Bragg reflection and the Bragg reflection takes place throughout the PhC structure. As a result, the quasimodes do not form a complete set and their spatial profile depends on the driving frequency, i.e, mode profile is dispersive.

The breakdown of completeness for an open cavity shows a signature in driven oscillation. If the cavity is  driven at frequency  $\omega$, the field inside the cavity will be $\textbf{E}(\textbf{r},t)= \textbf{E}(\textbf{r},\omega)\exp(-i\omega t)$ where $ \textbf{E}(\textbf{r},\omega)\neq \textbf{E}(\textbf{r},\omega_0)$ . Alternatively speaking, when an open cavity is driven, the wave function of the mode is determined by the driving frequency not the free oscillation frequency. 

In CROWs, therefore, the coupling between neighboring cavities is not characterized by the overlap of the eigen modes of the single cavity $\textbf{E}(\textbf{r},\omega_0)$ ($\omega_0$ is the eigen frequency of the cavity) but of the dispersive modes $\textbf{E}(\textbf{r},\omega)$.

In Fig. 2 the dispersion of a quasimode is shown qualitatively. In a true eigen mode, the size of the mode profile should be always the same as the cavity is driven at different frequencies. However, the quasimode is dispersive and will be spatial narrower (Fig. 2(a)) or wider (Fig. 2(b)) than the resonant mode when the driving frequency is lower or higher than the intrinsic frequency respectively. 
\begin{figure}[htbp]
\centerline{\includegraphics[width=.99\columnwidth]{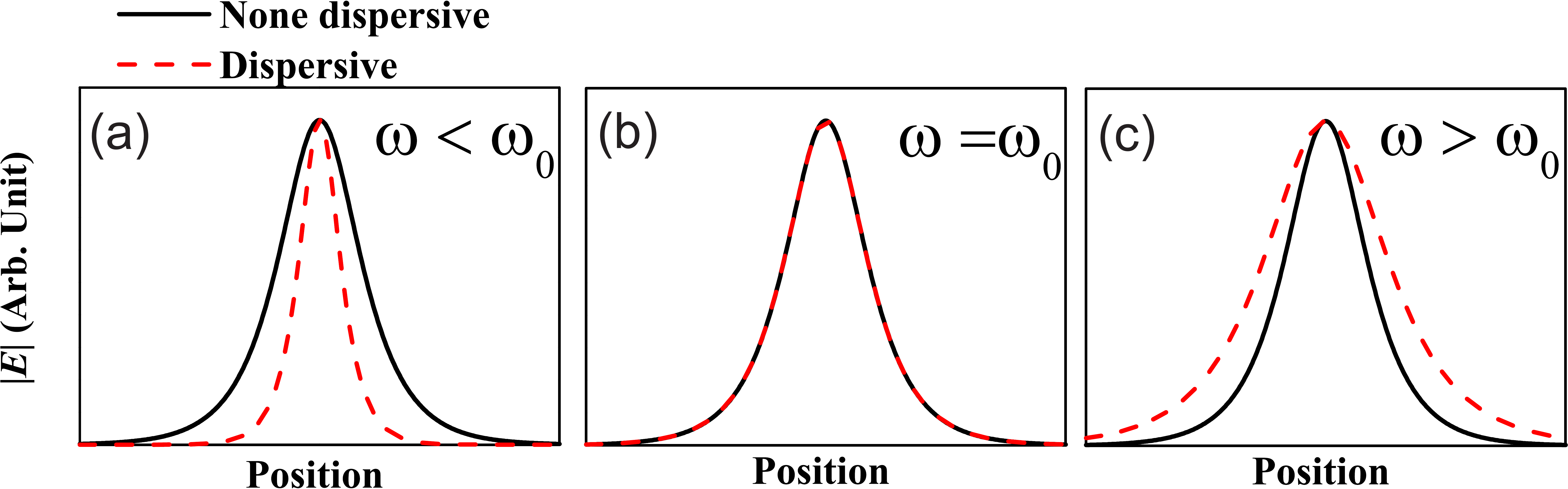}}
\caption{In driven oscillation, the quasimode (red) will be smaller (a) or larger (c) compared to the intrinsic resonant mode in the case when driving frequency is lower or higher than the intrinsic frequency. In contrast a true eigenmode (black) keeps the same spatial profile when driven off-resonance.}
\end{figure}

We performed two dimensional (2D) FDTD simulations to quantitatively study the dispersion of the mode profile of PhC mode-gap cavities. In the simulation a continuous source is placed in the center of a single cavity and switched on smoothly. After the transient died out, we output the $E_y$ field across the waveguide direction. The same procedure was repeated several times at different frequencies and different cavities \cite{Note1}.
\begin{figure}[htb]
\centerline{\includegraphics[width=.72\columnwidth]{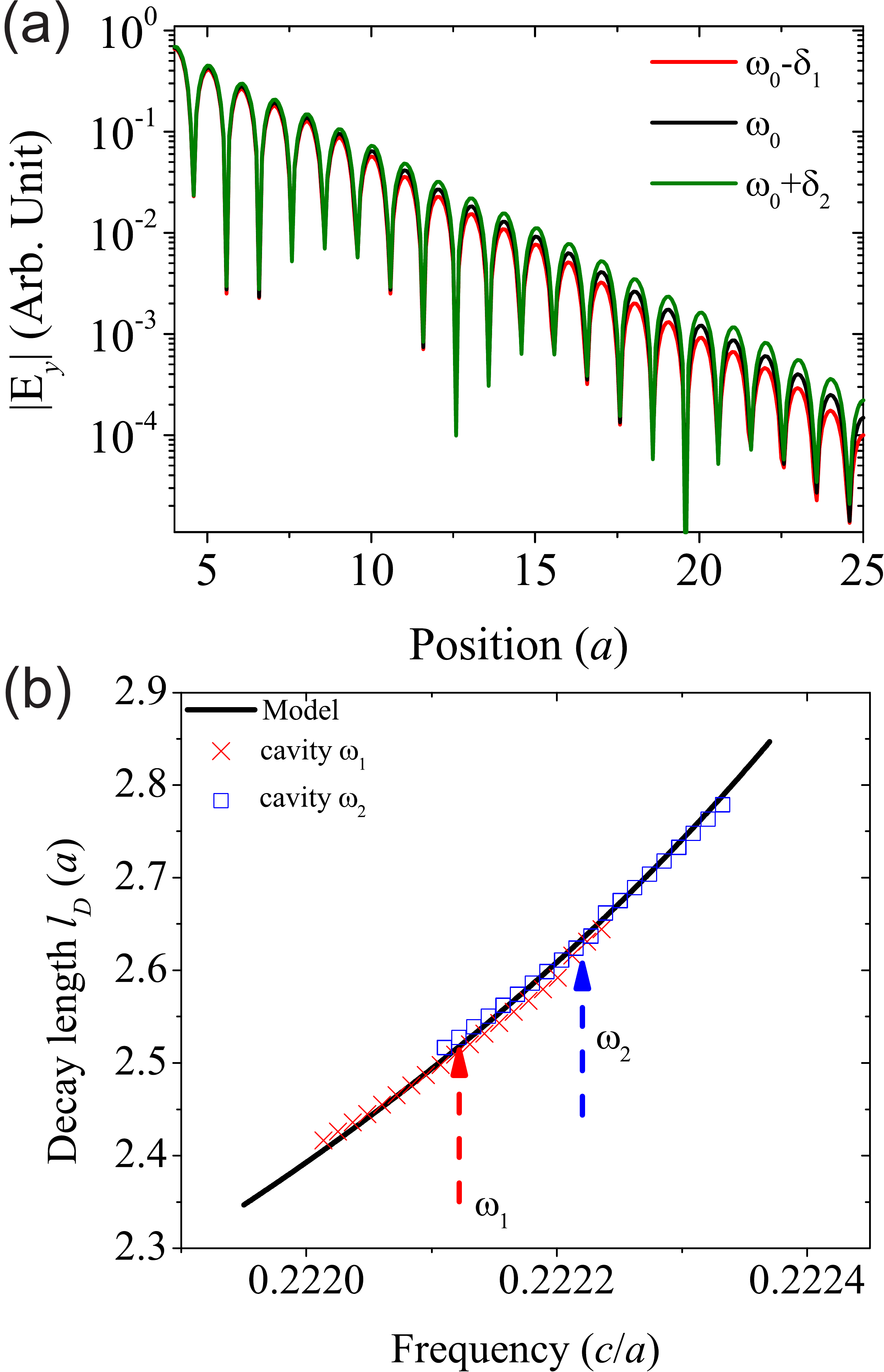}}
\caption{(a) 2D FDTD simulation of forced driving oscillation of a mode-gap cavity. The amplitude of the $E_y$ field across the waveguide direction versus the position is plotted. The center position is located at 0. The red, black and green lines illustrate cases with different driving frequencies ($\omega_0$=0.22212 $c/a$, $\delta_1$=0.00011 $c/a$ and $\delta_2$=0.00010 $c/a$).  (b) Red crosses (blue squares) depict decay lengths of a cavity with intrinsic frequency $\omega_1$=0.22212 $c/a$ ($\omega_2$=0.22222 $c/a$). The black curve is the model we use to describe the decay length. The decay lengths are obtained by fitting the envelope of the field from $x=4 a$ to $x=14 a$. The only difference between the geometries of the two cavities is the modulation strength.}
\end{figure}

In Fig. 3(a) we plot the amplitudes of the $E_y$ field versus the position across the waveguide direction when we drive the cavity at three different frequencies. When the driving frequency is $\omega_0+\delta_1$  which is higher than the resonant frequency $\omega_0$, the decay in space is slower. Alternatively speaking, the mode is larger in space. When the driving frequency is $\omega_0-\delta_2 $, the mode is smaller. This clearly shows that for a single mode-gap cavity, the mode profile differs depending on the driving frequency. Outside the modulation part of the cavity, the envelope of the mode profile decays exponentially. Along the waveguide direction $x$, the envelope decays as $\exp(-x/l_D)$, where $l_D$ is the decay length and $x=0$ is the center of the cavity. The decay length is related to the dispersion of the waveguide where the cavity lies in. The waveguide dispersion above the bandedge can be expanded as $\omega=\omega_{\textrm{edge}}+(k-\pi/a)^2/(2m)$ \cite{Baron2015}, where $m=(\partial^2\omega / \partial^2 k)^{-1}$ is the photon mass in analogy to the effective mass of electrons \citep{Ashcroft} and $\omega_{\textrm{edge}}$ is the frequency of the edge of the waveguide band.  By analytic continuation of the dispersion, the wave vector $k$ becomes a complex number when $\omega < \omega_{\textrm{edge}}$ . The decay length which is the inverse of the imaginary part of the wave vector,  follows as $l_D=(2m(\omega_{\textrm{edge}}-\omega))^{-1/2}$. In Fig. 3(b), we plot the calculated decay lengths by the analytical model and the extracted decay lengths from FDTD simulations of two different cavities at various driving frequencies. Firstly we see when the cavity is driven at different frequencies the decay lengths differ. Secondly for two different cavities (one with intrinsic frequency $\omega_1$ and the other with intrinsic frequency $\omega_2$), when the driving frequencies are the same, the decay lengths are the same. The decay lengths are well described by the analytical model, and we conclude that mode-gap cavities have a mode profile that depends on the driving frequency not the resonant frequency.

As the consequence of the dispersive mode profile,  the mode functions in the expression of the coupling rate \cite{Yariv1999,Haus1991} should be the dispersive mode (DM) not the eigen mode. The expression of the coupling rate becomes
\begin{equation}
\Gamma(\omega)=\frac{{\omega}\int\delta\epsilon(\textbf{r}-\textbf{R})\textbf{E}_{\omega}(\textbf{r}-\textbf{R})\cdot\textbf{E}_{\omega}(\textbf{r})d\textbf{r}}{\int\epsilon(\textbf{r})\textbf{E}_{\omega}(\textbf{r})\cdot\textbf{E}_{\omega}(\textbf{r})d\textbf{r}}.
\end{equation} In Eq.(1), $\textbf{E}_{\omega}(\textbf{r})$ is the electric field of the wave function at frequency $\omega$, $\epsilon(\textbf{r})$ is the dielectric constant of a single cavity and $\delta\epsilon(\textbf{r}-\textbf{R})$ is the dielectric difference between one isolated cavity and two cavities with heart to heart distance $\textbf{R}$. This expression is identical to that derived by Haus $et$ $al.$\cite{Haus1991} and by Yariv $et$ $al.$ \citep{Yariv1999}, except for the replacement of the resonant mode profile $\textbf{E}_{\omega_0}(\textbf{r})$  by the dispersive mode profile $\textbf{E}_{\omega}(\textbf{r})$. The resulting dispersion of the CROW is,\begin{equation}
\omega=\omega_0+\Delta+\Gamma(\omega)\cos(kR),
\end{equation} where $\omega_0$ is the eigen frequency of the single resonators, $\Delta$ is the frequency difference between $\omega_0$  and the center of the CROW band.
\begin{figure}[htbp]
\centerline{\includegraphics[width=.8\columnwidth]{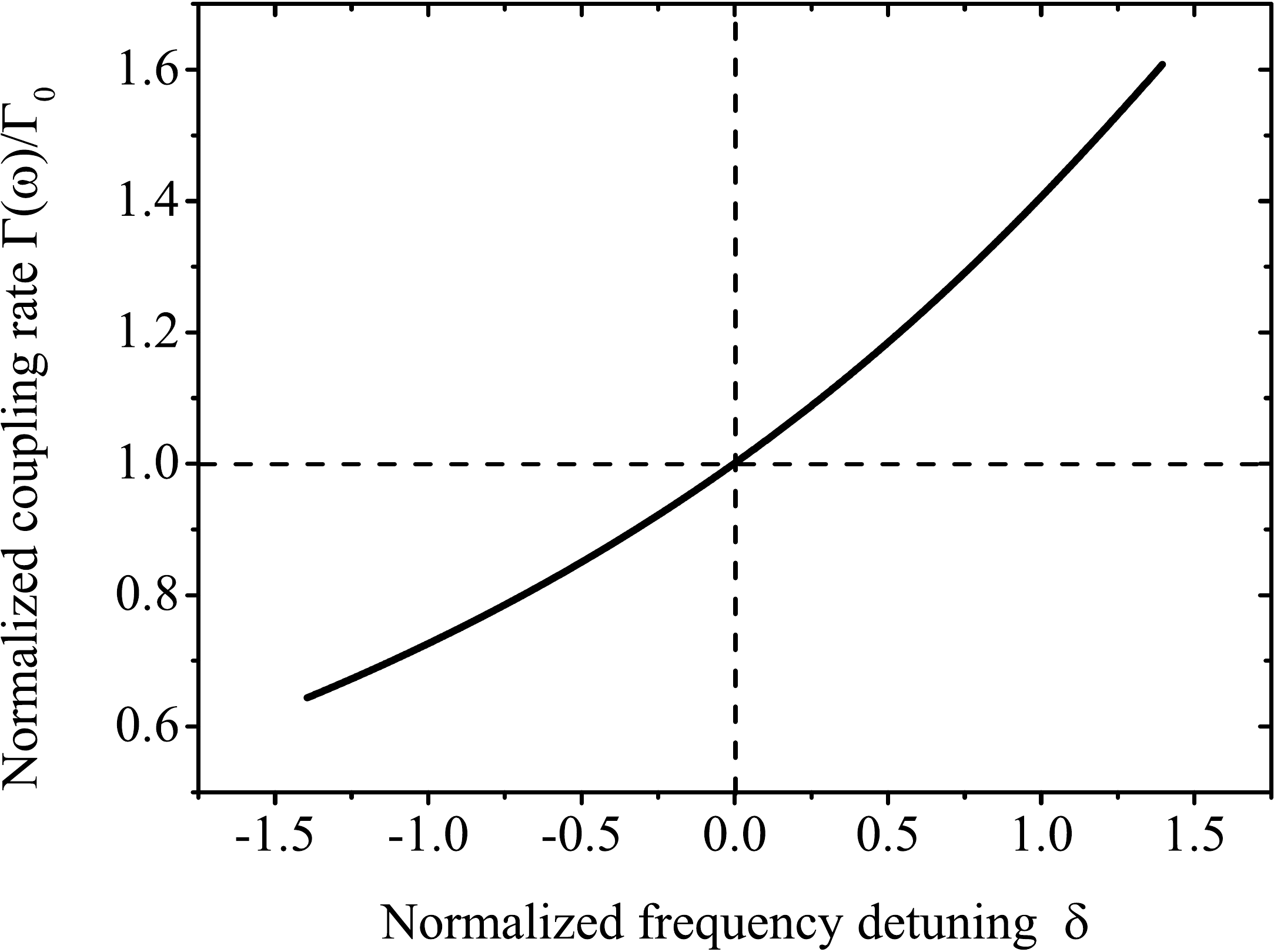}}
\caption{Normalized coupling rate  $\Gamma(\omega)/\Gamma_0$ of the CROW structure in Fig. 1 versus the normalized frequency detuning  $\delta$.}
\end{figure}

To evaluate Eq.(2) and (3),  some assumptions are made.  Firstly, we assume that each polarization of $\textbf{E}_{\omega}(\textbf{r})$ can be decomposed into $E^{X}_{\omega}(x)E^{YZ}_{\omega}(y,z)$, here $x$, $y$ represent the spatial coordinates and $x$ is along the waveguide direction, where the $E^{YZ}_{\omega}(yz)$ can be approximated to be non dispersive, $E^{YZ}_{\omega}(y,z)\approx E^{YZ}_{\omega_0}(y,z)$. Secondly, we approximate the envelop of $E^{X}_{\omega}(x)$ by $1/\cosh(x/l_D)$ , which is a resonable approximation for mode-gap cavities. The last assumption is for $\delta\epsilon(\textbf{r}-\textbf{R})$ and $\epsilon(\textbf{r})$. We divide $\delta\epsilon(\textbf{r}-\textbf{R})$ into 5 blocks. For each block, the effects of $\delta \epsilon$ is essentially the increase of the effective dielectric function of the waveguide. We approximate that this increase happens at the center of each block, so we express $\delta \epsilon$ as a collection of delta functions $\delta\epsilon=\sum\limits_{j=-2}^2 {\delta(x-x_j)\bar{\delta\epsilon}_j}$ with $x_j=ja $ and $\bar{\delta\epsilon}_j=\frac{3-|j|}{3}\bar{\delta\epsilon}_0$ (where $\bar{\delta\epsilon}_0$ is a constant). Based on these three assumptions the coupling rate is \begin{equation}
\Gamma(\omega)=\beta\sum\limits_{j=-2}^2 \frac{(3-|j|)\omega}{l_D\cosh{(\frac{R-ja}{l_D}}){\cosh(\frac{ja}{l_D})}}.
\end{equation} Here $\beta$ is a proportionality constant resulting from the integral over $y$, and we may abbreviate $\Gamma(\omega)=\beta g(\omega)$ where we emphasize $g(\omega)$ can be evaluated analytically. The dispersion relation becomes \begin{equation}
 \omega=\omega_0+\Delta+\beta g(\omega)\cos(kR),
\end{equation} there are only three unknowns ($\omega, \Delta, \beta$) which is exactly the same number of the unknowns the TB model has, i.e., our model does not introduce any extra free parameters.

We show the normalized coupling rate of the CROW structure in Fig. 1 as a function of normalized frequency detuning  $\delta=(\omega-\omega_0)/\Gamma_0$  with  $\Gamma_0=\Gamma(\omega_0)$ in  Fig. 4.  The normalized coupling rate is defined as $\Gamma(\omega)/\Gamma_0$, it is also equivalent to $g(\omega)/g(\omega_0)$. The normalized coupling rate  in Fig. 4 increases non-linearly as the detuning increases, in other words,  $\Gamma(\omega)$ is highly dispersive. At the low $k$ values which correspond to positive detuning $\delta$, the coupling rate is larger, and therefore the group velocity is enhanced. As a result, the maximum of the group velocity shifts to low $k$ values as compared to the result from the TB model.
\begin{figure}[htbp]
\centerline{\includegraphics[width=.72\columnwidth]{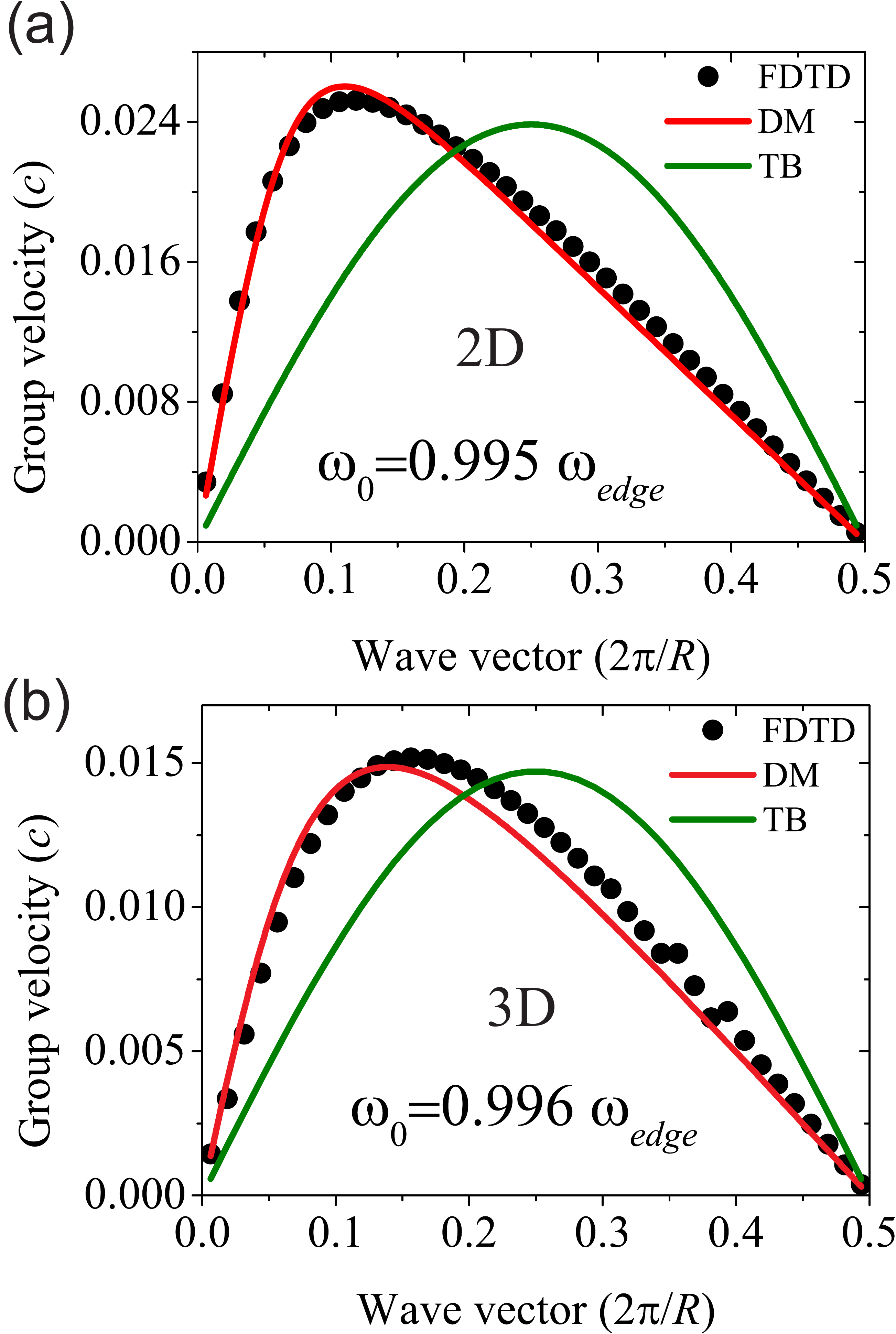}}
\caption{(a)  Group velocity of the 2D CROW consisting of mode gap cavities with modulation strength 1.5 calculated by FDTD (black dots), TB model (green line) and DM model (red line). All the parameters are as same as in simulations of driven oscillation. (b) Group velocity of the 3D CROW consisting of mode gap cavities with modulation strength 2 calculated by FDTD (black dots), TB model (green line) and DM model (red line). In the 3D simulation, $h=0.37a$ ($h$ is the thickness of the slab) and $\epsilon=10.04$.}
\end{figure}

We present the calculated group velocity of the CROW in 2D and in a 3D membrane structure in Fig. 5. In Fig. 5(a) the results of group velocity calculated by DM model is asymmetric with a maximum at $k \approx 0.10$ $(2\pi/R)$, this matches the 2D FDTD results extremely well. In Fig. 5(b) the group velocity from the DM model is asymmetric with maximum at $k \approx 0.13$ $(2\pi/R)$, it agrees satisfactorily with the fully 3D FDTD results \cite{Note2}. In contrast, the result from the TB model is symmetric about $k = 0.5$ $(\pi/R)$. The excellent match between our DM model and the FDTD data confirms that dispersion of the cavity quasimodes is indeed the physical reason for the failure of the TB model to accurately describe photonic crystal CROWs. 

In conclusion, the dispersion of a CROW composed of photonic crystal cavities can not be described by the standard TB mode due to the breakdown of completeness of the resonant modes. We show a new model taking into account of the dispersive property of the mode profile of the single cavity in a CROW structure. Our model describes the dispersion of the PhC CROW accurately without additional free parameters. Dispersion of the modes is inherent in all cavities. For racetrack resonators \citep{Xia2006,Xia2007} and PhC cavities with resonances far away from the bandedge such as L3 cavities \citep{Akahane2003}, dispersion is negligible and shows no discernible effect on the dispersion of the related CROW structures. However, for shallow defect cavities, such as mode-gap cavities with small modulation strength and double-heterostructure cavities \citep{Song2005, Jagerska2009} with small lattice mismatch, the defect modes are close to the edge of the waveguide band which make the dispersive nature of the modes become pronounced. The dispersion of CROWs of such cavities has strong asymmetry.  Thus, our theory will be useful for describing all devices based on coupling of shallow defect cavities such as delay lines \citep{Notomi2008},  optical memory \citep{Kuramochi2014}, and $\mathcal{PT}$ symmetric diodes \citep{Ramezani2014}.

The authors would like to thank Henri Thyrrestrup, Ad Lagendijk and Willem L. Vos for helpful discussions. This work was supported by European Research Council (project No. 279248), NWO-nano and NWO-Vici.

\bigskip
\noindent 


\end{document}